\documentclass[10pt,a4paper,twocolumn]{article}

\usepackage{lineno}
\usepackage{authblk}
\usepackage{booktabs}
\usepackage{graphicx}
\usepackage{tabularx}
\usepackage{makecell}
\usepackage[explicit]{titlesec}
\usepackage[labelfont=bf,labelsep=endash,font=footnotesize]{caption}
\usepackage{tabu}
\usepackage[inline]{enumitem}
\usepackage{xcolor}
\usepackage{csquotes}
\usepackage[
    backend=biber,
    natbib=true,
    style=numeric,
    maxnames=50,
    sorting=none
]{biblatex}
\usepackage[hidelinks]{hyperref}
\usepackage[hang]{footmisc}
\usepackage[normalem]{ulem}
\usepackage[top=2.5cm, bottom=2.8cm, left=1.5cm, right=1.5cm]{geometry}
\usepackage{mathtools}
\usepackage{balance}
\usepackage{wrapfig}
\usepackage{url}

\newcolumntype{b}{X}
\newcolumntype{m}{>{\hsize=.35\hsize}X}
\newcolumntype{g}{>{\hsize=.30\hsize}X}
\newcolumntype{s}{>{\hsize=.25\hsize}X}
\newcolumntype{t}{>{\hsize=.10\hsize}X}

\makeatletter
\renewcommand\AB@affilsepx{, \protect\Affilfont}
\makeatother

\providecommand{\keywords}[1]{\textbf{Keywords}\ \ \textendash\ \   #1}

\titleformat{\section}{\large\bfseries}{\thesection.}{1em}{\MakeUppercase{#1}}
\titlespacing*{\section}{0pt}{12pt}{6pt}

\titleformat{\subsection}{\large}{\thesubsection}{1em}{#1}
\titlespacing*{\subsection}{0pt}{12pt}{6pt}

\titleformat{\subsubsection}{\large\itshape}{\thesubsubsection}{1em}{#1}
\titlespacing*{\subsubsection}{0pt}{12pt}{6pt}

\newcommand{\ITUurl}[1]{\textcolor{blue}{\urlstyle{same}\url{#1}}}

\newcommand{\ITUpar}{\vspace{8pt}\par}

\newcommand{\ITUfootnote}[1]{\footnote{#1}}

\def\starttable{\vspace{6pt}\begin{table}[ht]\center}
\def\startfigure{\vspace{6pt}\begin{figure}[ht]\center}

\makeatletter
\def\tagform@#1{\maketag@@@{\ignorespaces#1\unskip\@@italiccorr}}
\makeatother

\title{\Large{\textbf{\uppercase{AI-Driven Container Security Approaches for 5G and Beyond: \\ A Survey}}}}

\author[1]{\normalsize{Ilter Taha Aktolga}}
\author[2]{\normalsize{Elif Sena Kuru}}
\author[3]{\normalsize{Yigit Sever}}
\author[4]{\normalsize{Pelin Angin}}

\affil[1,2,3,4]{\normalsize{Middle East Technical Univerity, Turkey}}

\date{\vspace{-12pt}{\small Corresponding author: Yigit Sever, yigit@ceng.metu.edu.tr} \\
\endgraf\rule{\textwidth}{1pt}}

\begin{document}

% \linenumbers %% this should be commented out if line numbers in the text are not wanted

\twocolumn[

\begin{@twocolumnfalse}
\maketitle

\begin{abstract}

The rising use of microservices based software deployment on the cloud leverages containerized software extensively.
The security of applications running inside containers as well as the container environment itself are critical infrastructure in the cloud setting and 5G.
To address the security concerns, research efforts have been focused on container security with subfields such as intrusion detection, malware detection and container placement strategies.
These security efforts are roughly divided into two categories: rule based approaches and machine learning that can respond to novel threats.
In this study, we have surveyed the container security literature focusing on approaches that leverage machine learning to address security challenges.

\end{abstract}
%%======%%

\ITUpar
%%= keywords =%%
%%======%%

% \ITUnote{Title, abstract and keywords must be identical to the ones submitted electronically in EDAS \textendash\ Editor's Assistant. Use the command \texttt{\textbackslash ITUnote} to achieve the appropriate formatting.}
\ITUpar
\ITUpar

\end{@twocolumnfalse}
]

\section{Introduction}%
\label{sec:intro}

Containers are lightweight and portable abstractions that contain the binary of an application as well as the necessary and sufficient minimal dependencies to run them.
Using containers to deploy software on the cloud has replaced the bare metal installations as the industry standard~\cite{vaucherSGXAware2018} due to microservices based architecture's demand for scalable and lightweight computation environments.
Companies such as Amazon, Netflix, Spotify and Twitter are using microservices architecture in their products~\cite{soldaniPains2018}.
Compared to the other contemporary isolation mechanism that is virtual machines, containers are faster to initialize and more lightweight since they do not need an extra virtualization layer to operate~\cite{kaurContainerasaService2017}.

The widespread adoption of 5G networks has led to an increase in the use of container technology to support the deployment and management of applications.
Containers are a straightforward answer for running the services required by 5G, they are portable and lean in terms of size requirements and lightweight in terms of preparation and startup times.
The use of containers in 5G networks can provide a number of benefits for Virtual Network Functions (VNFs), including improved scalability, flexibility, and efficiency.
Furthermore, the portability and flexibility of containers enable them to be deployed on-demand, making it easier to manage the lifecycle of VNFs and to adapt to changing network conditions.
Additionally, the use of container orchestration platforms such as Kubernetes allow for automated scaling and management of containerized VNFs, further improving the agility and scalability of 5G networks.
This makes it easy to deploy and run VNFs on any infrastructure, and to scale them up or down as needed.
Furthering the security and reliability of containers will, in turn, allow their rapid adoption in 5G VNFs~\cite{soualhiaAutomated2022}.

With the increasing adoption of container technology, there is a growing concern about the security of containerized applications and networks.
Containers are found to be less secure than virtual machines which is a detriment to their adoption~\cite{sultanContainer2019}.
The use of containers can introduce new vulnerabilities and risks that need to be addressed to ensure the security and integrity of 5G networks.
In this study, we conducted a survey of container security literature and focused our attention specifically on machine learning approaches on container security.

Machine learning (ML) techniques for container security are important and developing.
They have been investigated in various fields of research and implementation.
For example, Nassif et al.~\cite{nassifMachine2021} conducted a systematic review that analyzes machine learning models for anomaly detection.
They reviewed ML models from four perspectives: the application of anomaly detection, the type of ML technique, the ML model accuracy estimation, and the type of anomaly detection, whether they are supervised, semi-supervised or unsupervised.
A review conducted by Mohan et al.~\cite{mohanLeveraging2022} focused on the applications of various ML and deep learning methods in the implementation of defensive deception.
They summarized the classification of several deception categories, new machine learning and deep learning techniques in defensive deception, including the models, common datasets, key contributions, and limitations.
Moreover, Zhong et al.~\cite{zhongMachine2022} introduced a taxonomy of the most common machine learning algorithms used in the field of container orchestration.
The authors presented ML-based container orchestration approaches, classified the orchestration methods, and demonstrated the evaluation of ML-based approaches used in recent years.
Also, the autohors discussed machine learning approaches for anomaly detection.
Another survey conducted by Wong et al.~\cite{wongThreat2021} is a systematic review of containers, covering vulnerabilities, threats, and existing mitigation strategies, to provide information on the landscape of containers.
The authors also discussed some machine learning methods, and the papers used ML techniques to improve container security.
The current survey is different from those described above in various aspects, such as: Artificial intelligence solutions are included, such as artificial neural networks, machine learning, and deep learning solutions.
Supervised, semi-supervised, and unsupervised detection models and used datasets are included.
Focused on container security solutions, namely, intrusion detection, malware detection, attack detection, anomaly detection, and inter-container security included.

\section{Preliminaries}%
\label{sec:prelim}

We will base our discussion with the most prevalent container architecture for academia as well as commercial space; Linux containers.
Containers leverage two important Linux kernel features: control groups (cgroups) and namespaces.
A namespace is a layer of abstraction that covers the processes inside that namespace.
Wrapped processes get a private and isolated view of system resources.
The processes inside the namespace are also isolated from the changes that happen to global resources, allowing developers to prepare environments for binaries to run with defaults that the binaries expect and not disturb execution flow.
There are different types of namespaces that correspond to different constrained views into system resources.
There are a total of 8 different namespace types which constrain the view of either the cgroup root directory, message queues, network devices, mount points, process ID space, clocks, user and group IDs and hostnames.

\begin{figure}[h]
    \includegraphics[width=\columnwidth]{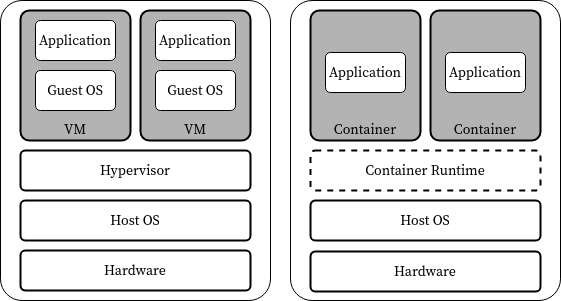}
    \caption{An overview of Virtual Machines and Containers}%
    \label{fig:vmvscontainer}
\end{figure}

As namespaces wrap processes with an isolated view of system resources, the level of allocation of said resources are controlled through cgroups.
cgroups are another Linux kernel feature which limits and monitors the resource usage of processes.
When a process is put into a cgroup hierarchy, it's access to system memory, CPU, priority of network communication, network bandwidth it can use etc. are controlled.
Control groups are used to allocate system resources fairly between different containers in the same host system.

All in all, containers are Linux processes that are constrained and isolated through aforementioned kernel features.
Through constraining and isolating the process or a bundle of related processes with files relevant to their operation, we get lightweight containers that can be packed with their dependencies.
Since setting up containers with cgroups and namespaces can get cumbersome, there are container management frameworks and container runtimes to assume these tasks.
Well known examples of these container technologies are Docker, Podman, Linux Containers (LXC), RKT and CRI-O.
A full-fledged cluster needs additional management and tooling as well.
These include service discovery within the cluster, container orchestration and networking among others~\cite{peinlDocker2016}.
Docker is favored as the main container technology~\cite{zouDocker2022} with thanks in no small part of Docker Hub, a public container library, playing a major role in its popularity.

Although the Linux kernel provides the ease of use for the isolation framework we have discussed, there is a drawback.
Since every container share the single kernel running in the host and there is no need for a separate hypervisor layer as in virtual machines, the isolation guarantees for containers are brittle.
Vulnerabilities and mismanaged containers can cause this isolation to be broken.
The result has been aptly titled \enquote{container escapes}~\cite{abbasPACED2022}.

Virtual machines are slower to start up and get running compared to containers.
One canonical question can arise at this point of the discussion: why do we use virtual machines if containers are more lightweight?
The security of containerized applications have been challenging researchers while the stronger isolation offered by virtual machines are better.
Furthermore, the kernel features that allow containers as we know them today have been matured much later than the framework to support virtual machines.

\subsection{Intrusion Detection Systems (IDS)}%
\label{sub:ids}

An intrusion in the context of computer security is attempted or successful access to confidential data or resources by unauthorized parties.
Network engineers use intrusion detection systems (IDS) which monitor a system and its resources to detect and report intrusions~\cite{shireyInternet2007}.
Monitoring system resources involves either placing sensors on host systems that analyze machine behavior or placing sensors on the network to monitor traffic.
IDSs are categorized according to these sensors: host-based based IDS (HIDS) and network-based IDS (NIDS), respectively.
Machine behavior that HIDS leverage can involve CPU, RAM and disk usage and network traffic that NIDS monitor involves individual packets that flow through the network and analytics that are derived from them~\cite{stallingsComputer2014}.

Another categorization we can apply to IDSs is whether they detect anomalous behavior by comparing against a set of predefined malicious behavior signatures or learning benign and malicious behavior to detect new behavior online.
The former is named signature-based IDS and the latter is named anomaly-based IDS.
Since developing signature-based IDS involves collecting and collating a large dataset which is not readily transferable from system to system~\cite{elkhairiContextualizing2022}, academia often focuses on anomaly-based IDS research.

\subsection{System Calls}%
\label{subsec:system-calls}

System calls are an interface between the hardware and the user space processes.
Processes interact with the kernel and request privileged actions, such as interacting with hardware resources or performing network operations.
These actions are restricted to certain processes and the kernel implements security policies to determine which processes can make certain system calls.
Since system calls are always present whenever a process performs a worthwhile action, it offers a valuable source of information.
Hence, system call monitoring is a common technique for detecting suspicious behavior in compromised applications because malicious code has to use system calls to perform malicious operations.
Tools like strace and ftrace are used to show the sequence of system calls made by a particular command or process~\cite{srinivasanProbabilistic2018}.
Monitoring system calls can help to identify and mitigate problems caused by compromised applications.

Bag of system calls (BoSC)~\cite{kangLearning2005} is a method for using system call data in machine learning applications.
The method involves creating a frequency list $S = \{s_{1}, s_{2}, \ldots, s_{n}$ where $s_i$ is the number of times the system call during that time window is observed~\cite{abedApplying2015}.
BoSC representation has seen frequent use in container intrusion detection literature~\cite{abedIntrusion2015, abedResilient2020} often paired with the Sysdig~\ITUfootnote{\url{https://sysdig.com/}} tool~\cite{rhlingStandardized2019, severEmpirical2022} to directly stream system calls from running containers with low overhead.

Frequency lists are not the sole method for using system call traces in machine learning applications
For instance, Srinivasan et al.~\cite{srinivasanProbabilistic2018} used sequence of system calls with preserved order to create $n$-grams with Maximum Likelihood Estimator for anomaly detection in containers.
Karn et al.~\cite{karnCryptomining2021} used n-gram representation as well during detecting malicious processes inside containers.
Iacovazzi and Raza~\cite{iacovazziEnsemble2022}, on the other hand, represented system calls in a sequence in a graph representation to preserve dependencies between system calls.
In a similar vein, Chen et al.~\cite{chenInformer2022} represented remote procedure calls with a graph to monitor microservice behaviour.

\section{Cyberattacks on Containers}%
\label{sec:cyber-attacks}

As previously mentioned, security concerns regarding containers are the major drawback against their adoption.
These security concerns have been categorized to lead the research community to study them on a common framework.
Sultan et al.~\cite{sultanContainer2019} investigated the threat model for containers across the literature and suggested 4 general use cases;
\begin{enumerate*}[label=(\roman*),itemjoin={,\;}]
    \item protecting containers from the applications inside
    \item protecting containers from each other
    \item protecting hosts from containers
    \item protecting containers from hosts
\end{enumerate*}
Tomar et al.~\cite{tomarDocker2020} extended these use cases by including the Docker client as a potential target.

For our discussion, we will handle the container security challenges from two perspectives.
First, the security of the application running in the container should be considered.
If an application has vulnerabilities or bugs, running it in an isolated setting will not prevent the loss of availability we will experience upon those vulnerabilities getting exploited.
Furthermore, we should mention the second area of interest before discussing the other consequences of application vulnerabilities.
We also need to consider the security of the containerization mechanism itself.
Securing the isolation and restriction of the runtime environment results in reliable systems.
Failure to do so can result in loss of confidentiality when containers operate in a multi-tenant environment through data leakages.
Another class of container vulnerability emerges when the isolation mechanism of the container is broken.
These vulnerabilities have been aptly named as \enquote{container escapes}.
Container escapes often abuse the interface offered to container development and runtime to access the host system.
In turn, those interfaces can be made accessible through the vulnerabilities in the applications themselves, allowing for arbitrary command execution inside the container environment.
Misconfiguration of containers or container runtime as well as the default privileges container runtimes have lead to privilege escalation in the host system and the eventual compromise of it.
The namespace feature of the kernel can also be exploited through namespace injection~\cite{leeKernelLevel2020}, which allows a malicious container to piggyback the hosts' isolation process and see the victim container's PID space just as the host can.
In this section, we will delve into one case of container escape in detail and analyze some attacks targeting the containerization process.
We will base our discussion around Common Vulnerabilities and Exposures (CVE), a public effort for collecting and publishing software vulnerabilities.

% CVE-2018-15664 - https://nvd.nist.gov/vuln/detail/CVE-2018-15664

CVE-2018-15664 is a vulnerability which leads to a container escape where the attacker gains free read-write access in the host system with root privileges.
The vulnerable API regarding this flaw in the docker engine is the \texttt{docker cp} call, which leverages FollowSymlinkInScope function which allows developers to resolve paths in containers.
However, Docker versions from 17.06.0-ce through 18.06.1-ce-rc2 suffer from a time-of-check to time-of-use vulnerability in FollowSymlinkInScope.
Since the resolution step of the path and actually using the path are not performed sequentially, there exists a time frame where the attackers can symlink a resolved path to an arbitrary place, which includes root owned directories in the host machine~\cite{saraiOsssec2019}.

% CVE-2019-5736 - https://nvd.nist.gov/vuln/detail/CVE-2019-5736

CVE-2019-5736 is a vulnerability that stems from the runc binary up to version 1.0.0:rc6.
runc is a container runtime that Docker as well as CRI-O, containerd and Kubernetes uses.
The flaw in effected runc binary versions allow an attacker to use a malicious container to overwrite the runc binary in the host system and gain root access and privileges~\cite{saraiOsssecurity2019}.
The only prerequisite the vulnerability requires is any command to be run as root from the container where said container creates a new container using an attacker controlled image or running \texttt{docker exec} to get a shell from an already running container which gave the attacker write access previously.
Prior to proper patching, this vulnerability could be prevented by using namespaces correctly and mapping the root of the host system and the container's user into different namespaces.

\section{Machine Learning Approaches for Container Security}%
\label{sec:ml-for-contsec}

Container security is handled through rule based matching utilities where known vulnerabilities and common misconfiguration errors are collated through human effort~\cite{minnaSecurity2022}.
These utilities are adequate for catching known attacks and configuration mistakes developers make.
However, they cannot detect attacks or vulnerabilities missing from their rule set.
To tackle this issue, machine learning based container security solutions have been developed.
In this section, we will survey the container security approaches that leverage machine learning.

\subsection{Intrusion Detection}%
\label{sub:intrusion-detection}

Zhang et al.~\cite{zhangRealtime2021} proposed an intrusion detection system for Digital Data Marketplace (DDM).
The presented system utilizes One-Class Support Vector Machine (OC-SVM) algorithm.
The OC-SVM algorithm is an unsupervised learning method that finds a decision boundary with maximum distance from data points, making it suitable for anomaly detection where training data is unbalanced.
Similar to SVM's hyperplane, it uses spherical boundary to separate data.
They capture system calls using fixed size window, apply preprocessing and then feed into the ML model.
Besides intrusion detection, they match output of the detection module with attack database to decide whether the anomaly is linked to other anomalies.
Their dataset contains system call data from database applications and machine learning applications which are running in containers.
For database containers, they have used Sysdig.
They generated traffic with Apache JMeter\ITUfootnote{\url{https://jmeter.apache.org/}}.
In addition, for unusual traffic, Metasploit for Nmap is used.
For machine learning containers, they have also used Sysdig to detect adversarial attacks during the training. The trained model successfully detected 100\% of the arbitrary code executions and brute force attacks with a low false positive rate. They state ROC curve values reach up to 0.995. Machine learning containers, all the models except TPGD, attacks detected with 100\% This work is limited with system calls does not consider any other criterion.

\begin{table*}[!ht]
    \begin{center}
        \caption{Overview of surveyed intrusion detection in container approaches}%
        \label{tab:intrusion-detection}
        \begin{small}
            \begin{tabularx}{\linewidth}{@{} t g m m m b s @{}}
                \toprule
                \thead{Work} & \thead{ML\\ Method} & \thead{Feature\\ Collection} & \thead{Dataset} & \thead{Victim\\ Machine} & \thead{Attack\\ Type} & \thead{Monitoring}\\
                \midrule
                \cite{zhangRealtime2021} & OC-SVM & n-gram & custom syscall attack dataset & CoughDB, MongoDB, static ML app & Container Escalation, Brute Force Execute Arbitrary Code, Adversarial ML attacks & JMeter, nmap, Sysdig \\
                \cite{elkhairiContextualizing2022} & auto-encoder & system call sequence graph & LID-DS, CB-DS & Flask-python web app & Sprocket Information Leak, MySQL Auth Bypass, Release Agent Abuse, Dirty Pipe & Sysdig \\
                \cite{severEmpirical2022} & REPTree, Random Tree, Random Forest, SMO & BoSC, network flow & custom syscall and network flow dataset & rConfig & OS Command Injection, SQL Command Injection & Sysdig, tcpdump \\
                \cite{floraUsing2020} & STIDE, BoSC, HMM classifiers & STIDE, BoSC & custom syscall dataset & MariaDB & Overflow, Bypass, Privilege Escalation, DoS & Sysdig \\
                \cite{cavalcantiPerformance2021} & Decision Tree, Random Forest & BoSC & custom syscall dataset & MySQL & Authentication Bypass, DoS, Privilege Escalation, Integer Overflow & Sysdig \\
                \cite{iacovazziEnsemble2022} & Random Forest, Isolation Forest & anonymous walk embedding & custom syscall dataset and CUI-2020 & Hadoop cluster, NGINX, Apache Solr & Cryptomining, backdoor & perf \\
                \cite{popeContainer2021} & semi-supervised learning & process graph, node2vec & auditd & \emph{contribution} & DoS, privilege escalation & auditd \\
                \cite{wangContainerGuard2022} & variational autoencoder & time series performance event data & Container Performance Event Dataset (CEPD) & Container based big data platform & Spectre, Meltdown & ptrace, perf \\
                \cite{chakravarthiDeep2022} & auto-encoder, GAN & network traffic, system and network level performance data & VM Migration dataset created using CloudSim & 2 host with 3 VM in total  %% TODO how to say it, is it ok?
                & Net Scan, DoS & Not mentioned  \\ %% TODO couldnt find tool
                \cite{shenClustered2022} & Random Forest & n-gram & ADFA-LD & Django, Httpd, MySQL, Tomcat & XSS Attack, SQL Injection, Security Policy Bypass, Remote Command Injection, Identity Bypass, Arbitrary File Read/Write & Sysdig \\
                \bottomrule
            \end{tabularx}
        \end{small}
    \end{center}
\end{table*}

El Khairi et al.~\cite{elkhairiContextualizing2022} proposed a HIDS that relies on monitoring system calls.
The authors used Sysdig to collect the system calls.
The novelty of their work comes from their usage of context information alongside system calls to build a graph structure to train and test their IDS.
Context information includes system call arguments and recently observed system calls.
The authors report that they were motivated to use context information due to the shortcomings of existing HIDS approaches.
They used LID-DS dataset~\cite{grimmerModern2019} and extended the dataset using their contributed dataset: CB-DS, which consists of container escapes.

Here, we will explain their feature selection in detail.
First, they built a graph representation of the system calls with argument information.
This graph representation natively includes the recently seen system calls as well.
A graph constructed for a timeframe $t$ under benign conditions can be then used in the set of normal behavior expected during container's normal operation.
When the training is over and testing begins, any unseen vector is classified against the previously constructed benign dataset of graphs.
The authors evaluated their framework on different classes of vulnerabilities and compared their approach against CDL and STIDE-BoSC.

Sever et al.~\cite{severEmpirical2022} tackled a research gap in container IDS literature.
The authors realized that previous work focused solely on HIDS approaches that monitor system calls to train and evaluate anomaly-based IDS.
This leaves anomaly-based NIDS which can leverage network traffic features such as network flow out of the picture.
In order to answer whether this omission is justified or not, the authors set up an experiment environment with JMeter as the benign traffic source, a web application running in a container as the victim and the Metasploit tool as the malicious traffic source.
The authors used Sysdig tool to gather system call traces and tcpdump to capture network traffic between the attacker machine and the victim container.
They used system call data with BoSC as the feature and network flow data derived from network \texttt{.pcap} captures.
In order to evaluate intrusion detection performance, the authors selected 4 machine learning algorithms found commonly in the container IDS literature: REPTree, random tree, random forest and SMO.
After evaluating both BoSC and network flow based monitoring with those 4 algorithms, the authors found that network flow data yielded better performance than BoSC.
However, the authors used only one victim application with only 3 different attacks for their dataset, putting a detriment on the generalizability of their study.

Flora et al.~\cite{floraUsing2020} evaluated intrusion detection performance by monitoring system calls on a containerized application by using attack injection.
Their approach to this comparison is twofold: they evaluated the instruction detection performance between Docker containers, LXC and the application running on bare metal while using three classifiers: BoSC, STIDE, and HMM.
First, the authors decided on an application: MariaDB, running in a container for the Docker and LXC settings and standalone for the bare metal case.
As is the case with attack injection approaches, they decided on the TPC-C workload for the benign traffic source.
For malicious traffic, they picked 5 CVEs and used their implementations from \url{exploit-db.com}.
The authors decided to capture every system call emitted by the containers using sysdig tool during the experiments while they captured only MariaDB and it's children's system calls for the bare metal case.
Running the TPC-C workload for 24 hours with 30 minutes of malicious traffic during benign traffic period yielded the data required for the analysis.
The authors then used to classifiers to discern between malicious and benign traffic.

During their analysis, the authors found that intrusion detection by using the methods mentioned above yielded the best overall results for the application running in the Docker container.
While detection on Docker gave the highest recall across all three algorithms, BoSC performed marginally better than STIDE and wholly better than HMM.
The authors also concluded that using lower epochs resulted in better detection performance and interpreted it as the models learning how to discern between the malicious and the benign traffic without learning unnecessary details.
On the other hand, the author's analysis is constrained on only one database application: MariaDB.

Cavalcanti et al.~\cite{cavalcantiPerformance2021} compared the performance of intrusion detection systems for containers.
They framed their observations under two categories: the effect of the classifier architecture and the performance of different machine learning algorithms.
The authors set up an attack injection scenario where they subjected a MySQL Docker image to TPC-C benchmark for benign traffic and 4 different attacks from \url{exploit-db.com} with CVEs for malicious traffic.

Overall, the authors used three classifier architectures: label encoding and one-hot encoding, sliding window with label encoding and one-hot encoding, and sliding window with BoSC.
All in all, the authors used AdaBoost, Decision Tree, Gaussian Naive Bayes, K-Nearest Neighbors, Multi-layer Perceptron, Multinominal Naive Bayes, Random Forest and Support Vector Machine.
Gaussian Naive Bayes performed the best in terms of recall while K-Nearest Neighbors had the best precision out of all machine learning algorithms for the first classifier architecture.
In terms of F-Measure, Support Vector Machine had the highest performance with 83.2\%.
Due to the number of algorithms and the measures involved, we will continue our discussion constrained to F-Measure.
The second classifier architecture achieved the highest F-Measure of 99.4\% with the Random Forest algorithm when the window size was 30.
Both decision tree and random forest had the highest F-Measure with 99.8\% for the final classifier architecture when the sliding window size was set to 30 again, albeit not much higher than other algorithms.
The important takeaway from the results obtained by the authors is that the context of which the system calls appear contributes more to the detection performance than the specific machine learning algorithm chosen.

Iacovazzi and Raza~\cite{iacovazziEnsemble2022} present a machine learning-based solution for intrusion detection in cloud containers.
The proposed solution combines supervised and unsupervised learning methods, and it is designed to work at the host operating system level, using data observable at the kernel level.
The solution uses a mix of random forests and isolation forests to classify container workload behaviors and detect adverse behavior within the containers.
Note that random forests are supervised learning methods while isolation forest are unsupervised.
First, a graph representation of the sequence of system calls are collected at the host machine's kernel level.
This graph is then processed using random walks and anonymous walks algorithms to extract the features.
This representation is fed into a random forest classifier, which is trained on normal classes and outputs a set of probabilities for whether the input belongs to each class.
The probabilities are passed to a third stage, where they are used for generating anomaly scores using an ensemble of isolation forest modules, one for each normal class.
Isolation forest modules are trained on datasets containing samples from the respective normal class and contaminated with samples from other normal classes.
The final decision about the class of the input sample is based on the outcomes of the anomaly scores.
If all anomaly scores are below a threshold, the input is classified as the class with the highest score or as an anomaly if all scores are under the threshold.
If more than one score is higher than the threshold, the input is classified as an anomaly.
In order to effectively capture dependencies among adjacent system calls in a sequence, which are not considered in the bag-of-system-calls approach, they use a graph-based representation.
This graph representation and feature extraction process enables the effective classification of container workload behaviors and the detection of malicious behavior within the containers. % graph vs BoSC comparison %
Despite the results of the EoF method outperforms SVM and LOF alternatives, there were some limitations to this approach, as it was not able to detect all attacks with a true positive rate above 0.7 namely Backdoor and SQL Injection.
Moreover, the work has been done on Docker containers and possible attacks during container migration haven't been discussed.

In their work, Pope et al.~\cite{popeContainer2021} introduce a new dataset derived from the Linux Auditing System, which contains both malicious and benign examples of container activity.
This dataset is the first of its kind to focus on kernel-based container escapes and includes attacks such as denial of service and privilege escalation.
The data was generated using the autoCES framework and includes partial labels identifying benign and malicious system calls over specific time intervals.
However, the dataset has some limitations, including incomplete annotations and a limited number of container escape scenarios.
Additionally, the selection of benign background activity in the dataset may not be comprehensive.
The goal of this dataset is to be used in a semi-supervised machine learning context.
For the machine learning process, they began by converting the auditd data into a process graph, which illustrated the relationships between processes.
This graph was then transformed into vectors using a node embedding technique.
The resulting vectors were used to train a logistic regression classifier, which was able to accurately predict whether a process was benign or malicious with a F1 score of 97\%.
The authors also mention that the dataset could potentially be utilized for other applications, such as training an autoencoder for anomaly detection.
These results demonstrate the effectiveness of the dataset in a semi-supervised learning context.

%% Another autoencoder based IDS
Another work by Wang et al.~\cite{wangContainerGuard2022} propose a real-time intrusion detection system.
They focus on detecting meltdown and Spectre attacks in the container environments.
Spectre and Meltdown are vulnerabilities that can be exploited using
%% should we mention what is cache based side channel attack?
cache-based side-channel attacks to access sensitive data.
These vulnerabilities allow attackers to access data that is temporarily stored in the cache, which can then be extracted using cache-based side-channel attacks.
In this work, to satisfy conditions for Spectre and Meltdown attacks, the scenario is designed as the containers were co-resident (i.e. sharing the same hardware).
They designed the ContainerGuard service to watch the workflows.
By monitoring, they capture hardware and software performance time-series data.
After data collection, they distribute data to corresponding variational autoencoders considering the performance data category which are hardware CPU events, hardware cache events and software events.
For the purpose of evaluating a method for detecting the Meltdown and Spectre attacks, a dataset called the container performance event dataset which includes 400,000 benign and 60,000 malicious data was created.
The method's highest AUC score ranges from 0.90 to 0.99.
In addition to the detection performance, there is no significant runtime performance overhead which is measured as approximately 4.5\%.

%% And also another autoencoder based IDS too
Chakravarthi et al.~\cite{chakravarthiDeep2022} focus is on assessing the effectiveness of anomaly detection during service and virtual migrations in the cloud environments.

%% The commented paragraph below is moved to the dataset section
%% They used the CloudSim 5.0 environment and collected traffic data.
%% They augmented the data using a generative adversarial network %% (GAN)%%

The authors trained the autoencoder and SVM on the generated dataset.
The performance of two different classifiers, an autoencoder and a support vector machine, were evaluated using ROC curves.

They state that autoencoder performs well during VM migrations with a false positive rate below 15\%.
They used the reconstruction error of the AE model as the anomaly score.
One limitation of their work is that there is no benchmarked dataset available to test the resilience of cloud infrastructure.
They have generated data samples from a simulated network and balanced them using the GAN network.
These samples have been classified as either anomalous or normal using the AE model.
However, their trained model is only able to detect anomalous traffic in a cloud environment that is similar to the one simulated in their experiments.

The clustering algorithms aim to divide the provided unlabeled data into clusters that achieve high inner similarity and outer dissimilarity.
They do not rely on signatures, a description of attack classes, or labeled data, therefore;
for the purpose of detecting anomalies in unlabeled data, unsupervised IDS and clustering approaches are used.

To increase the effectiveness of anomaly detection in the edge computing environment, Shen et al.~\cite{shenClustered2022} suggested an anomaly detection framework combining cluster algorithms.
The proposed framework initially identifies and classifies containers before building anomaly detection for each group.
Also, they use system calls to inspect containers' behavior and perform classification and intrusion detection.
They looked into eight real-world vulnerabilities, and the experiment result shows that the framework increased the True Positive Rate (TPR) from $90.3\%$ to $96.2\%$, and False Positive Rate (FPR) reduced from $0.61\%$ to $0.09\%$ compared to the traditional method.

The framework utilizes Sysdig to collect system call data generated by containers, the DBSCAN cluster algorithm to classify containers in an unsupervised way, and the RandomForest classifier for each application category to detect anomalies.
Also, they use their approach against two different detection methods.
First, they use one detector for all containers.
This method collects system calls from all applications without distinguishing the application.
The other method is using one detector for each container.
Even though the second approach achieves better results, it incurs a remarkable performance cost.

\subsection{Malware Detection}%
\label{sub:malware-detection}

% Everything I want to say is here but maybe be a bit nicer?
Wang et al.~\cite{wangDockerWatch2022} designed and implemented a malware detection framework for containerized applications.
The novelty of their work comes from their approach to extract executables from containers with respect to the container's storage driver type.
The authors decided to support overlay2 and aufs, the current and past recommended storage drivers respectively.
With the executable in hand, the suggested framework first uses disassembled code and binary itself for fast path coarse detection using a multichannel CNN.
The slow path detection is done using a LSTM-CNN with API-call sequences as the features.

The authors have evaluated their implementation on 3000 malware samples acquired from VirusShare and 300 container specific attacks against 2000 benign binaries.
Even though the authors compared their framework against previous work under metrics such as precision and recall, the previous work they opted to compare to are not from container security domain but deal in general software security.
Hence, the 300 container specific attacks are mixed in with the rest of the 3000 malware samples and there is no particular insight presented for regular software shipped in containers and vulnerable containers.

% could be named better
Cryptomining malware has become a significant threat in Kubernetes, with hidden executables that uses server resources for mining.
To detect and classify pods that hold cryptomining processes, Karn et al.~\cite{karnCryptomining2021} proposed that machine learning can be used together with system calls.
They used several types of cryptominer images, namely Bitcoin, Bytecoin, Vertcoin, Dashcoin, and Litecoin.
Also, they included healthy pods, that are MySQL, Cassandra, Hadoop, Graph, Analytics and Deeplearning.
They captured system calls with a period of 1 minute for each pod.
Then they leveraged n-grams to extract features.
After numerous experiments they decided to set n as 35 due to its high recall rate.
Following the feature extraction, four ML models which are decision tree, ensemble learning, feed-forward vanilla artificial neural network and feedback recurrent neural network have been selected to train with the data collected.
% Results of each ML model %%
The accuracy of ensemble learning model from Python-XgBoost library was similar on training and validation sets, 89.3\% and 89.4\% respectively.
For feed-forward Vanilla ANN, they have used the combination of Keras and Tensorflow, with autokeras tool to tune hyperparameters.
Overall performance was 81.1\% on training set, and 79.7 \% on validation.
Due to the nature of the system calls, it is suitable to use it as time-series data.
Therefore, they implemented LSTM RNN.
The accuracy on the training set was 79.99 and  78.90\% on validation set.
Decision tree implementation with default parameter values using python's SKLearn library achieved 99.6\% accuracy on training and 97.1\% on validation by beating all other models.
% text below may be unnecessary %
In addition, for better model explainability and visual representation, they have used SHAP and LIME tools.

\subsection{Attack Detection}%
\label{sub:attack-detection}

Lin et al.~\cite{linSHIL2022} proposed an attack detection framework which consists of different layers in a pipeline in an attempt to increase detection rate while addressing false positive and lack of labelled training data issues.
Their proposal has 3 different modules; first, they employ an unsupervised anomaly detection layer which uses an autoencoder neural network.
The authors claim that the encoder and the subsequent decoder will generate results with a high reconstruction error for anomalous samples.
The second layer in the pipeline uses random forest algorithm to cross validate edge cases and potentially eliminate false positives.
On the final layer, the authors employ an isolation forest, a self-supervised model in order to detect outliers and generate training labels automatically.
This pipeline is fed with system call frequency vectors, acquired using Sysdig with a sampling rate of 100 milliseconds.

In order to evaluate their proposed framework, the authors applied 7 minutes worth of benign traffic onto the containers using JMeter, where applicable.
At the start of the 5th minute, the authors started the attack, some attacks caused the container to crash which ended the experiment but for the rest, the attack completed and the experiment ran until the JMeter is finished at the 7th minute.
Lin et al. compared their proposed framework against CDL~\cite{linCDL2020}, self-patch, a supervised random forest approach and a supervised CNN.
They used 41 real world attacks with assigned CVEs, encompassing 28 applications.
They used containerized applications with application vulnerabilities, not container specific attacks.

Lin et al.~\cite{linCDL2020} presented a classified distributed learning framework, namely CDL, to detect anomalies in containerized applications. The framework achieves anomaly detection in four major steps: System Call Feature Extraction, Application Classification, System Call Data Grouping, Classified Learning, and Detection. They process the raw system call trace into a stream of frequency vectors, and these extracted feature vectors are used to identify applications. For the identification of applications, they utilize random forest learning scheme ~\cite{hoRandom1995}.
Random forest classifier uses numerous decision trees, then chooses the most voted result among individual decision trees. Hence, the random forest model gives the predicted application classification result. When this process has identified the containers of the same application, the framework makes a system call data grouping to append the frequency vector traces of different containers and use them for model training and attack detection. Lastly, for anomaly detection, the unsupervised model uses autoencoder neural networks.
The authors investigated 33 real-world vulnerabilities documented in Common Vulnerabilities and Exposures  (CVE) database, and the results show that CDL can detect 31 out of 33 attacks. Also, they inspected the system run time, and the data indicates that CDL is lightweight and suitable for detecting attacks in real time under real-world circumstances.

\subsection{Anomaly Detection}%
\label{sub:anomaly-detection}

Gantikow et al.~\cite{gantikowContainer2020} investigated the behavior of containers by using neural networks to detect anomalies.
The authors present two approaches for anomaly detection based on system call traces.
First, system call distributions are used to detect anomalies.
One layer Long Short Term Memory (LSTM) network is trained to predict the system call distribution at time $t+1$ based on distribution at time $t$.
%% The evaluation demonstrates that this approach is ineffective for detecting mimicry attacks when an anomaly makes a few system calls while the application uses many simultaneously.
The second approach is a neural network using file/directory paths for anomaly detection.
Their method is based on training a neural network to predict the following file system path based on the most recent file system path used by a system call.
The proposed neural network consists of a Word Embedding Layer, followed by LSTM layers which are designed to learn to predict the following file system path based on the vector representation of the current one.
After a prediction was made by this neural network, the actual file path and predicted path were compared to detect anomalies.

\begin{table*}[ht]
    \begin{center}
        \caption{Overview of surveyed anomaly detection in container approaches}%
        \label{tab:anomaly-detection}
        \begin{small}
            \begin{tabularx}{\linewidth}{@{} t b b b @{}}
                \toprule
                \thead{Work} & \thead{ML\\Model} & \thead{Data\\Used} & \thead{Collecting\\Method} \\
                \midrule
                \cite{gantikowContainer2020}  & LSTM                            & system call, file/directory path                                                               & Sysdig \\
                \cite{wangUnsupervised2022}   & BiLSTM                          & system call                                                                                    & ptrace \\
                \cite{castanhelTaking2021}    & KNN, RF, MLP, AB                & system call                                                                                    & strace \\
                \cite{cuiUnsupervised2021}    & LSTM auto-encoder               & system call                                                                                    & Sysdig \\
                \cite{tunde-onadeleStudy2019} & KNN, k-means SOM                & system call                                                                                    & CoreOS clair, Sysdig, JMeter \\
                \cite{duAnomaly2018}              & KNN, SVM, NB, RF                & performance monitoring data                                                                    & cAdvisor, Heapster \\
                \cite{chenInformer2022}       & DCRNN                           & RPC traffic                                                                                    & RPC chain clustering \\
                \cite{kamthaniaNovel2019}     & Restricted Boltzman Mahcine     & user and system defined security profile, automated NIST violations, run-time security profile & python script \\
                \cite{tienKubAnomaly2019}     & \emph{contributed}              & system call, Network, I/O activities                                                           & JMeter, Sysdig \\
                \cite{ulhaqueKGSecConfig2022} & LR, NB, SVM, RF, XGB            & security related config documents                                                              & BeautifulSoup, NLTK \\
                \cite{kosinskaDetection2022}  & SARIMA, HMM, LSTM, auto-encoder & system metrics (streaming data)                                                                & Prometheus \\
                \bottomrule
            \end{tabularx}
        \end{small}
    \end{center}
\end{table*}

Wang et al.~\cite{wangUnsupervised2022} proposed an unsupervised anomaly detection framework.
The authors initially acquired system call sequences from the ptrace tool.
Then, they used word2vec technique to map each system call within their context from the sequences into a fixed size vector.
These vectors are used sequentially for the rest of the author's instruction detection framework.
A BiLSTM variational auto encoder is used.
At the final layer of their framework, the authors detected anomalies through reconstruction error.
For evaluation, the authors employed the UNM system call sequences dataset.
They have also extended it with system call sequences gathered during a sqlmap attack on a container running MySQL as well as 3 different container escape attacks.
The dataset they used for evaluation consisted of 0.63\% anomalous traces with benign samples as the rest.
Overall, their approach yielded 90\% accuracy and an F1 score of 90.75\%.

Castanhel et al.~\cite{castanhelTaking2021} presents an approach for using system calls to detect anomalies in containerized systems.
The authors focus on how the size of the window impacts the results through the implementation of a sliding window technique.
In the paper, the authors first discuss the challenges of monitoring containers and the importance of detecting anomalies in order to ensure the security and stability.
In their implementation, they collected a dataset of system calls by running strace on the host machine, outside the container, from a variety of containerized applications and used machine learning techniques to train a model to classify normal and anomalous system calls based on this dataset.

The dataset used in the study consisted of 50 traces of system calls, with half representing normal behavior and the other half representing anomalous behavior.
The normal behavior traces consisted of five different types of expected interactions with the WordPress application, while the anomalous behavior traces consisted of five different types of attacks focused on cross-site scripting (XSS) and remote code execution (RCE).

The experiments in the study were conducted on a Linux host using Docker, and the application used for testing was WordPress, popular open-source content management system that is served by the Apache web server.
The collected system calls were divided into four groups, with the first group containing the most dangerous system calls that alter system behavior.
The last group contained harmless system calls that primarily query to get system behavior rather than issuing commands.

A sliding window technique was used to analyze data from various sources and four algorithms (KNN, RF, MLP, and AB) were applied using seven different window sizes.
The data was split into training and testing sets and ten executions were run for each classifier to evaluate the results and prevent overfitting.
The random seed was changed during the split phase to generate different sets.

They tested both with all data and the data without harmless system calls and found that the model was able to accurately detect anomalies in the system calls of containerized applications, with an average accuracy of over 90\%.
Overall, the paper concludes that system calls can be an effective means of detecting anomalies in containerized systems but also mentions the fact that their work does not contain all calls available in current systems.
Also, as mentioned, by completing tasks using a variety of containers with different applications instead of just the WordPress application with additional plugins, the dataset would have been more diverse, allowing for a more comprehensive evaluation of the system's stability.

% it is also a dataset paper.
% cites other dataset related works for IDS

%% Paragraphs below are moved to the dataset section.

Cui and Umphress~\cite{cuiUnsupervised2021} come up with open-source dataset for the observation of system calls.
%% The current work aims to improve upon previous datasets used for detecting anomalies in computer systems by addressing some of their limitations.
%% These include focusing on network traces rather than internal system behaviors, limited scope and coverage in system call-based datasets, and a lack of clear descriptions of benign behaviors and indications of system activity.
%% Additionally, none of the previous datasets are explicitly designed for containerized systems.
%% In response, the current work has been created using Docker.
%% Brute force login, simple remote shell, malicious python script, SQL misbehavior, SQL injection, docker escape and other selected malware were used.
They have used sliding window with a fixed size.
They have chosen to use a classic Long Short-Term Memory (LSTM) autoencoder as the baseline classifier for the unsupervised classification task.

Autoencoders are a type of neural network model that were first introduced by Rumelhart et al.~\cite{rumelhartLearning1986}
for unsupervised learning of compact representations, or encoding of input data.
These models consist of two main parts: an encoder, which maps the input data to a lower-dimensional latent space, and a decoder, which maps the latent representation back to the original input space.
The encoder and decoder are trained together by minimizing a reconstruction loss function that measures the difference between the original input and the reconstructed output.

The idea of using Long Short-Term Memory (LSTM) units in autoencoder architectures likely emerged after the introduction of LSTMs by Hochreiter and Schmidhuber
for modeling sequential data, and the development of autoencoders for feature learning and dimensionality reduction.

The reason behind LSTM selection is that it has the ability to remember and use knowledge from previous batches which makes it suitable for anomaly detection.
In this experiment, a total of 42 models were trained using different combinations of configurations, including 7 different window sizes, 3 different feature sets, and 2 normalization methods.
These models were then tested on 7 different attacks, with 6 different confidence levels applied, resulting in a total of 1764 entries.
Overall model predicts with over 90\% accuracy for brute force login, meterpreter, malicious script and remote shell attacks.
However, it's accuracy on docker escape attacks was only 76.27\%.
Moreover, it was observed that the proposed framework was only evaluated using an offline dataset and a single application, and it is planned to conduct a comprehensive online evaluation.
Besides, while the current work has successfully demonstrated the potential for unsupervised introspection, it is necessary to expand the dataset to include multiple applications to see its potential in different contexts.

%maybe we can talk about static and dynamic detection schemes, difference etc. before this paragraph <elif>
Tunde et al.~\cite{tunde-onadeleStudy2019} presented a combination of static and dynamic anomaly detection schemes to detect security vulnerabilities for containers.
They conducted a study on static and dynamic vulnerability detection strategies using 28 common real-world security vulnerabilities discovered in Docker Hub images.
Firstly, they used CoreOS Clair, an open-source static analysis engine that scans containers layer-by-layer for known vulnerabilities using Common Vulnerabilities and Exposures (CVE) databases.
Afterward, they investigate dynamic detection schemes using different unsupervised machine learning algorithms.
These machine learning algorithms are selected to address the following unique challenges of container security:

\begin{enumerate}
    \item Containers are short-lived, so the detection algorithms can not use large amounts of training data.
    \item Containers are highly dynamic; thus,  the detection algorithms cannot make any assumptions about the application or attack behavior in advance.
    \item The detection algorithms should be able to detect vulnerabilities with low overhead.
\end{enumerate}

Properties above of container exploit detection lead to using light-weight unsupervised anomaly detection schemes such as K-Nearest Neighbor (k-NN) Algorithm, K-Means Clustering, KNN combined with Principal Component Analysis (PCA), and Self-Organizing Map (SOM).
Their comparison between different exploit detection schemes was based on four metrics: detection coverage, false positive rate, and lead time.
The metrics indicate if each approach can detect vulnerabilities, how accurately they can achieve detection and how quickly they can detect attacks, respectively.
The k-NN algorithm is used to perform outlier detection.
Because the presence of noise in the feature data prevents the k-NN algorithm from achieving high- accuracy, they used k-NN with PCA.
While the k-NN algorithm can detect $32.14\%$, PCA + k-NN succeeds in a slightly better detection with $35.71\%$.
The k-means approach achieves a $67.86\%$ detection coverage rate.
The Self-Organizing Map (SOM) approach over system call time vectors (SOM time) detects $75\%$ of vulnerabilities, while the SOM approach over system call frequency vectors (SOM frequency) detects $79\%$ of vulnerabilities.
Therefore, the SOM approach accomplishes the highest detection coverage.
At the false positive rate comparison, again SOM approach achieves the lowest false positive rate, with $1.7\%$ for SOM frequency and $1.9\%$ for the SOM time.
It is followed by the K-means clustering approach with $7.67\%$.
Moreover, k-NN and k-NN with PCA obtain the highest false positive rates with $9.92\%$ and $9.88\%$, respectively.
Finally, the SOM approach attained the largest detection lead time, with 28.7 for SOM frequency and 25.8 for SOM time.
Nevertheless, k-NN, k-NN with PCA, and K-means achieve 0.57, 1, and 0.36 seconds respectively.
Furthermore, the paper states that combining static and dynamic schemes can increase the detection coverage rate to $86\%$.
In conclusion, the authors show that static analysis for container security is insufficient.
In contrast, using unsupervised machine learning algorithms, dynamic anomaly detection schemes can succeed high detection coverage rate with a low false positive rate.
The study demonstrates that Self Organizing Map algorithm is better than the other mentioned algorithms in terms of all three metrics.

Du et al.~\cite{duAnomaly2018} used different supervised machine-learning algorithms to detect and diagnose anomalies in container-based microservices.
They proposed an anomaly detection system (ADS) by analyzing real-time performance data for anomaly detection and diagnosis.
The proposed ADS consist of three modules: the monitoring module, the data processing module, and the fault injection module.
First, the monitoring module is used to collect real-time performance monitoring data from the target system.
In this paper, the authors focused on container and microservice monitoring, and the term \enquote{container} was used to refer to a collection of containers constituting one complete microservice.
Secondly, the data processing module is used to analyze this data and detect anomalies.
They determine whether a container performs well by gathering and processing its performance data, just as they determine whether a microservice is abnormal by gathering and processing the performance data of all related containers.
After classifying if a microservice experience an anomaly, the ADS finds the anomalous container.
In order to detect anomalies, they use supervised machine-learning algorithms such as Support Vector Machines (SVM), Random Forest (RF), Naive Bayes (NB), and Nearest Neighbors (NN).
Also, to find the container that caused an anomaly, they used time-series analysis.
Lastly, the fault injection module simulates service faults (CPU consumption, memory leak, network package loss, and network latency increase) and collects datasets of performance monitoring data.
These datasets are used to train machine learning models to validate the anomaly detection performance.
In their experiments, three datasets are structured according to this module, and the three services they choose using Clearwater, an open-source virtual IP Multimedia Subsystem.

The validation results show that Random Forest and Nearest Neighbors classifier gives satisfying results using each dataset.
Furthermore, SVM performs the worst since it does not work well on datasets with multiple classes.
All in all, if the dataset is created using only one service, the authors recommend the NN classifier.

Remote procedure calls (RPC) allow components in a distributed cluster of applications to invoke each other's functions (procedures) seamlessly, as if those functions are owned by the invoking application.
The network layer between the components is abstracted away as a result.
Chen et al.~\cite{chenInformer2022} suggested using RPCs as an alternative to monitoring system calls, since RPCs are required for meaningful interaction between a distributed cluster's components just like how system calls are required for worthwhile operations within an application.
They handled RPCs as RPC chains, a sequence of RPCs that depend on each other and appear in order during common operations.
The authors found that representing RPC chains as directed weighted graphs suits their use case well.
They represented nodes as RPCs, edges and weights as the dependency between different RPCs, and labelled nodes with the number of times that particular RPC was invoked.

To learn regular RPC traffic and predict anomalous RPC traffic, the authors first trained DBSCAN~\cite{esterDensitybased1996} clustering algorithm to acquire the RPC chains.
The authors then trained a DCRNN model to predict the traffic model from previously observed RPC traffic.
By using mean absolute error and variants, they managed to label anomalous traffic when observed RPC chains deviated from the expected traffic in their case study which was performed on a Kubernetes cluster with \enquote{billions of daily active users}~\cite{chenInformer2022} and RPC traffic that spans 2 weeks.

Kamthania~\cite{kamthaniaNovel2019} presents a deep learning-based algorithm for detecting malicious patterns in individual container instances.
The algorithm is designed to be easily applied to any container platform that adheres to the Open Container Initiative (OCI) standard.
The algorithm utilizes a Gaussian-Bernoulli restricted boltzmann machine.

% what is restricted boltzmann machine?
A restricted boltzmann machine (RBM) is a type of neural network that is used for unsupervised learning.
RBMs are composed of a visible layer, which encodes the input data, and a hidden layer, which learns features from the input data.
RBMs are trained using an energy-based model, where the energy of a configuration of the visible and hidden units is minimized.
RBMs are often used for tasks such as dimensionality reduction and collaborative filtering.
Gaussian-Bernoulli RBMs are a variant of RBMs that can handle continuous-valued data, rather than just binary data, in the visible layer.

By using RBM, they create a container profile based on the configuration of the containers and extract behavioral statistics at runtime.
The algorithm then uses automated NIST container security rules to identify any security violations for the container under test and applies a machine learning algorithm to build a complete security profile for the container.
In their results they mention classification rate for some attack types: unbounded network access from containers, insecure runtime configurations, rogue containers, improper user access rights, embedded clear texts.
However, the details of the classification rate is not mentioned.

%% What is OC-SVM

KubAnomaly, a system that offers security monitoring capabilities for anomaly detection on the Kubernetes orchestration platform was suggested by Tien et al.~\cite{tienKubAnomaly2019}.
The aim of this system is to improve Docker security which is compatible with Kubernetes.
KubAnomaly provides a security-monitoring module with customized rules in Sysdig and observes the internal activities of containers, system calls, I/O activities, and network connections.
Since monitoring too many events would result in large overhead, they selected four system call categories which are file I/O, network I/O, scheduler and memory.
Furthermore, to identify hackers and insider intrusion events, it performs anomaly detection using machine learning classification.
A neural network model was created to classify multiple types of anomalous behavior, such as injection attacks and denial-of-service (DoS) attacks.
This machine learning model uses supervised learning and three different datasets had been used to train the model.
These three different datasets are: private data, a public data called CERT and real-world experimental data to evaluate the system accuracy and performance.
Further explanation about the datasets are in the Section~\ref{sub:datasets-container-security}

The proposed anomaly classification model is organized into four steps.
They begin by monitoring log data from their agent service, which collects monitor logs from Docker-based containers.
After obtaining the raw monitor logs, they extract features to train their models.
The next step is data normalization for fast convergence and improved accuracy.
To apply data normalization StandardScaler, MinMaxScaler, and Normalizer provided by the machine learning framework sklearn.
Lastly, they construct the anomaly classification model using four fully connected layers, and for the backend, they use Keras and Tensorflow.
For the purpose of apply the classification model to the real world they designed the KubAnomaly system.
In addition, they developed an online web service with vulnerabilities and tested the system.
The results show that KubAnomaly is able to detect many abnormal behaviors.

Mubin et al.~\cite{ulhaqueKGSecConfig2022} focuses on configurations of container orchestrators.
Container orchestrator itself should be correctly configured to provide security for all other managed containers.
They introduce a new method that uses keywords and learning to capture knowledge about configurations which was not studied before.
The module created, namely KGSecConfig aims to create a Knowledge Graph for Configuration (KGConfig) of various platforms, cloud providers, and tools used in CO to organize scattered data.
They extracted information from documentation files and created entities.
Between these entities, several relationships such as \enquote{hasDefault}, \enquote{hasArgument}, \enquote{hasType}, \enquote{hasOption},  and \enquote{hasDescription}.
This representation is used to identify the configuration syntax and formulate keyword-based rules for estimating the relevancy of security documents with configuration.

In order to train a supervised learning model to extract configuration concepts from documents, a labeled dataset was needed.
Since no labeled dataset existed, 3,300 sentences were labeled by two authors according to the four configuration concepts.
3,032 sentences that were agreed upon by both labellers were used for training the model to reduce labeling bias.
Five  machine learning classifiers, Logistic Regression (LR), Naive Bayesian (NB), Support Vector Machines (SVM), Random Forest (RF), and Extreme Gradient Boosting (XGB) were selected for the learning-based models, and various features such as TF-IDF-based word level, character level, and combination of word and character level, NLP features, were considered.
The optimal traditional ML models were selected using Bayesian optimization and average Matthews Correction Coefficient with early stopping criteria.
Breadth-First-Search algorithm was used to identify the configuration argument and update the KGConfig.
Accuracies of the LR, NB, SVM, RF, XGB were 0.94, 0.82, 0.88, 0.76, 0.93 respectively.
The model's results showed that KGSecConfig is effective in automating the mitigation of misconfigurations.

Kosinska and Tobiasz~\cite{kosinskaDetection2022} proposed a system, namely, Kubernetes Anomaly Detector (KAD), for detecting an anomaly in a Kubernetes cluster.
KAD uses various machine learning models to achieve high accuracy.
Their solution differs from other solutions in using different machine learning models that facilitate detecting different types of anomalies.
The KAD system chooses the appropriate model for detection;
thus, different models can be matched to different data types.
These models are SARIMA, HMM, LSTM and Autoencoder.
SARIMA and HMM are derived from traditional time series and statistical models;
LSTM and HMM are deep learning models.
In their experiment, they trained models the models trained on the Numenta Anomaly Benchmark (NAB) dataset.
They selected two types of data streams: the first stream is artificially generated, and the second contains data presenting CPU utilization collected from AWS Cloudwatch.
The results show that statistical models (SARIMA and HMM) achieve higher results on the artificial data and the LSTM and autoencoder perform better on AWS Cloudwatch data.
Furthermore, the experiments demonstrate that the real-time anomaly detection capabilities of the KAD system can be successfully deployed in a Kubernetes cluster.
However, the KAD system allows anomaly detection to be performed on one metric at a time.
Hence, for more complex cases, multivariate models can be needed.

\subsection{Inter-container Security}%
\label{sub:inter-container-security}

Deng et al.~\cite{dengSecure2022} tackled the secure placement of containers on cloud setting where multiple residents share the same host.
The motivation for their work comes from the author's findings that container placement strategies do not consider the security of co-resident containers.
As mentioned in Section~\ref{sec:cyber-attacks}, container vulnerabilities also pose risks to the container running on the same host as well as the host system.
Deng et al. considered the whole placement challenge as a series of placement decisions.
This allowed the authors to reframe the problem as an optimization task.
The authors then used deep reinforcement learning (DRL) with the encoded placement policy as the input.
DRL is a branch of machine learning that uses reinforcement learning and deep learning principles.
Nguyen et al.~\cite{nguyenDeep2021} state that DRL is suited to solve complex, dynamic, and high-dimensional cyber-defense problems.
Hence, DRL is used for container security.
Deng et al's model output the placement decision for the given container at every time step and the reward function is a formula with a trade-off between security and workload balancing.
With their evaluation, the authors find their model to perform better than existing strategies in terms of workload balance while keeping security in mind.

Li et al.~\cite{liOptimal2022} proposed a defensive deception framework for container-based clouds.
Their approach generates an adversarial model, decoy placement strategy, and decoy routing tables using a DRL Algorithm.
First, they developed an adversarial model, namely the System Risk Graph (SRG).
SRG extracts risks and threats in the container-based cloud and includes overall risks and vulnerabilities from the application to the visualization layer.
Secondly, $SRG_{t}$, which is the system risk graph of the cloud at time slot-$t$, is sent to input neurons of the DRL agent.
DRL Algorithm generates an ideal decoy placement strategy to decide optimal topologically locations and types on digital decoys assets.
Moreover, the performance of the placement strategy is used as the reward data to train and update the DRL agent.
This feature enables the DRL agent to evolve with the dynamic cloud.
Therefore, their method is adaptive and fully interacts with the dynamic environment.
Lastly, the determined placement strategy and deceptive routing for the decoy are sent to the orchestration platform.
As a result, the proposed framework increases the detection ratio on the random-walker attack by $30.69\%$ and the persistent attack by $51.10\%$.

Using genetic algorithm, Kong et al.~\cite{kongSecure2019} suggested a Secure Container Deployment Strategy (SecCDS) to defend against co-resident attacks in container clouds.
SecCDS reduce co-residency by 50\% compared with existing strategies by coordinating the placement and migration of containers to separate attackers and victims on different physical machines (PMs).
In their paper, they define two metrics to describe the deployment and co-residency of container clouds.
Deployment Matrix (DM) represents the correspondence between container and PM, and Coresidency Matrix (CM) describes co-residency among different tenants in the cloud.
Later, they develop a deployment strategy by Genetic Algorithm that can detect relational aggression among different tenants in real-time and dynamically migrate containers, effectively preventing co-residency.
In this implementation, a container must only be deployed in a unique position and belong to a real tenant.
Thus, the authors offered a genetic mechanism with altered crossover and mutation operations of traditional genetic algorithm by changing some mutation operation steps and proposed a new individual learning mechanism.
In addition, they utilize Simulated Annealing (SA) to do a neighborhood search for each individual in GA, which helps the algorithm reach the optimal global solution.

\subsection{Dataset for Container Security}%
\label{sub:datasets-container-security}

Chakravarthi et al.~\cite{chakravarthiDeep2022} used the CloudSim 5.0 environment and collected traffic data.
They augmented the data using a generative adversarial network (GAN).
The dataset contains TCP traffic.
Since the work focuses on anomaly detection in scenarios where VM migration occurs, to detect particularly volume-based attacks, by analyzing the traces of TCP streams is beneficial since these types of attacks tend to consume an excessive amount of bandwidth compared to normal traffic.
The collected data contains wide variety of features, some of which are CPU, network and memory usages.
In addition, they gathered information about the status of the network flow.
To solve data imbalance problem, the authors decided to augment data based on the collected samples.
For this task, they selected GAN~\cite{chakravarthiDeep2022} rather than restricted boltzman machines or variational autoencoders.
Creating new samples becomes a challenging task when the data has many variables or features.
During simulation, they have introduced Net Scan and DoS attacks, but the proposed work does not include information about the statistical properties of the dataset, such as the percentage of data collected for each metric, which would be useful in understanding the characteristics of the data.

Cui and Umphress~\cite{cuiUnsupervised2021} come up with open-source dataset for the observation of system calls.
Scripts for data generation are available in their public repository.
The work aims to improve upon previous datasets used for detecting anomalies in computer systems by addressing some of their limitations.
These include focusing on network traces rather than internal system behaviors, limited scope and coverage in system call-based datasets, and a lack of clear descriptions of benign behaviors and indications of system activity.
Additionally, at that time, the authors state that none of the previous datasets are explicitly designed for containerized systems.
In response, the current work has been created using Docker.
Brute force login, simple remote shell, malicious python script, SQL misbehavior, SQL injection, docker escape and other selected malware types were included in the dataset.
Besides, they captured 7,144,780 benign system calls which constitutes the majority of the dataset.
As a limitation, the sample application used in this study was MySQL database.
Therefore, scenarios during the experiment may not be identical to all real-world environments.

Tien et al.~\cite{tienKubAnomaly2019} use three different datasets to train the supervised machine learning model for the security system called KubAnomaly.
These datasets are private data, public data called CERT, and real-world experimental data to evaluate the system's accuracy and performance.
First, they used a simple dataset and a complex dataset.
These datasets contain two parts: 80\% for training and 20\% for testing, and both datasets include normal and abnormal samples.
Normal samples include several types of web services run in containers.
They used JMeter to simulate user login behavior.
The abnormal samples include two types of attacks aimed at compromising the web service.
They used Owasp Zap to simulate a hacker's attempt to attack the container and JMeter to simulate a DoS attack.
The complex dataset also contains both normal and abnormal sample types.
Also, the complex dataset includes other hacker tools such as sqlmap.
KubAnomaly achieves over 98\% accuracy with the simple dataset and 96\% accuracy with the complex dataset.
CERT contains various types of log data, inclusive of email and device data., but it does not include system call log data.
This dataset does not have any labeling;
therefore, they used feature extraction and unsupervised learning to classify abnormal user behavior.
Finally, in order to attract hackers, the authors developed an online web service with vulnerabilities and used KubAnomaly to identify abnormal behaviors and record the attack events.

In their experiment for the selection of anomaly detection model, Kosinska and Tobiasz~\cite{kosinskaDetection2022} used the Numenta Anomaly Benchmark (NAB) dataset to train the models.
They chose two different sorts of data streams: the first is artificially generated, and the second contains data presenting CPU utilization collected from AWS Cloudwatch.
The results show that statistical models (SARIMA and HMM) achieve higher results on the artificial data while the LSTM and autoencoder perform better on AWS Cloudwatch data.

\section{Conclusion}%
\label{sec:conclusion}

In conclusion, this survey has provided a comprehensive overview of container security in 5G environments and the potential of AI-based methods to address the challenges posed by increased connectivity.
The integration of AI into security systems has the potential to enhance intrusion and malware detection, anomaly detection, attack detection, and inter-container security within container clusters, making it a powerful tool in the fight against cyberattacks.
It is important to note that challenges such as interpretability, explainability, and bias need to be addressed when integrating AI in container security.
Nevertheless, the use of AI-based methods for container security in 5G environments has the potential to revolutionize the way we protect and secure our digital assets.
Further research and development in this field is needed to fully realize the potential of AI-based approaches for container security in 5G environments.
This survey aims to contribute to the growing body of knowledge in this field and provide a valuable resource for researchers, practitioners, and decision-makers working in container security and 5G networks.

\section*{Acknowledgement}
\label{sec:ackn}
This research has been supported by the TÜBİTAK 3501 Career Development Program under grant number 120E537. However, the entire responsibility of the publication belongs to the owners of the research. The financial support received from TÜBİTAK does not mean that the content of the publication is approved in a scientific sense by TÜBİTAK.\ITUpar

\printbibliography

\section*{Authors}%
\label{sec:auth}

\begin{wrapfigure}{l}{0.32\columnwidth}
    \vspace{-.1in}
    \includegraphics[width=0.39\columnwidth]{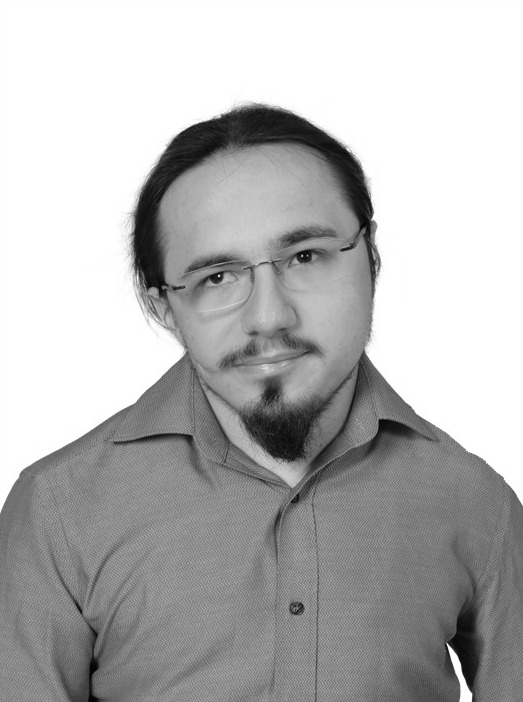}
\end{wrapfigure}\textbf{Ilter Taha Aktolga}
received his B.S. degree in computer engineering at Middle East Technical University (METU) in 2021. He is currently a graduate student pursuing a Master's degree at METU. He has gained research and industry experience through various positions, including working as an undergraduate researcher at KOVAN Robotics Research Laboratory at METU with a TUBITAK scholarship from June 2019 to June 2020 and as a cloud developer at Arcelik Global from May 2021 to August 2021. Currently, he is working as a software engineer at ASELSAN. His research interests include the fields of machine learning, container security, and backend development with an emphasis on microservices.\ITUpar

\begin{wrapfigure}{l}{0.32\columnwidth}
    \vspace{-.1in}
    \includegraphics[width=0.39\columnwidth]{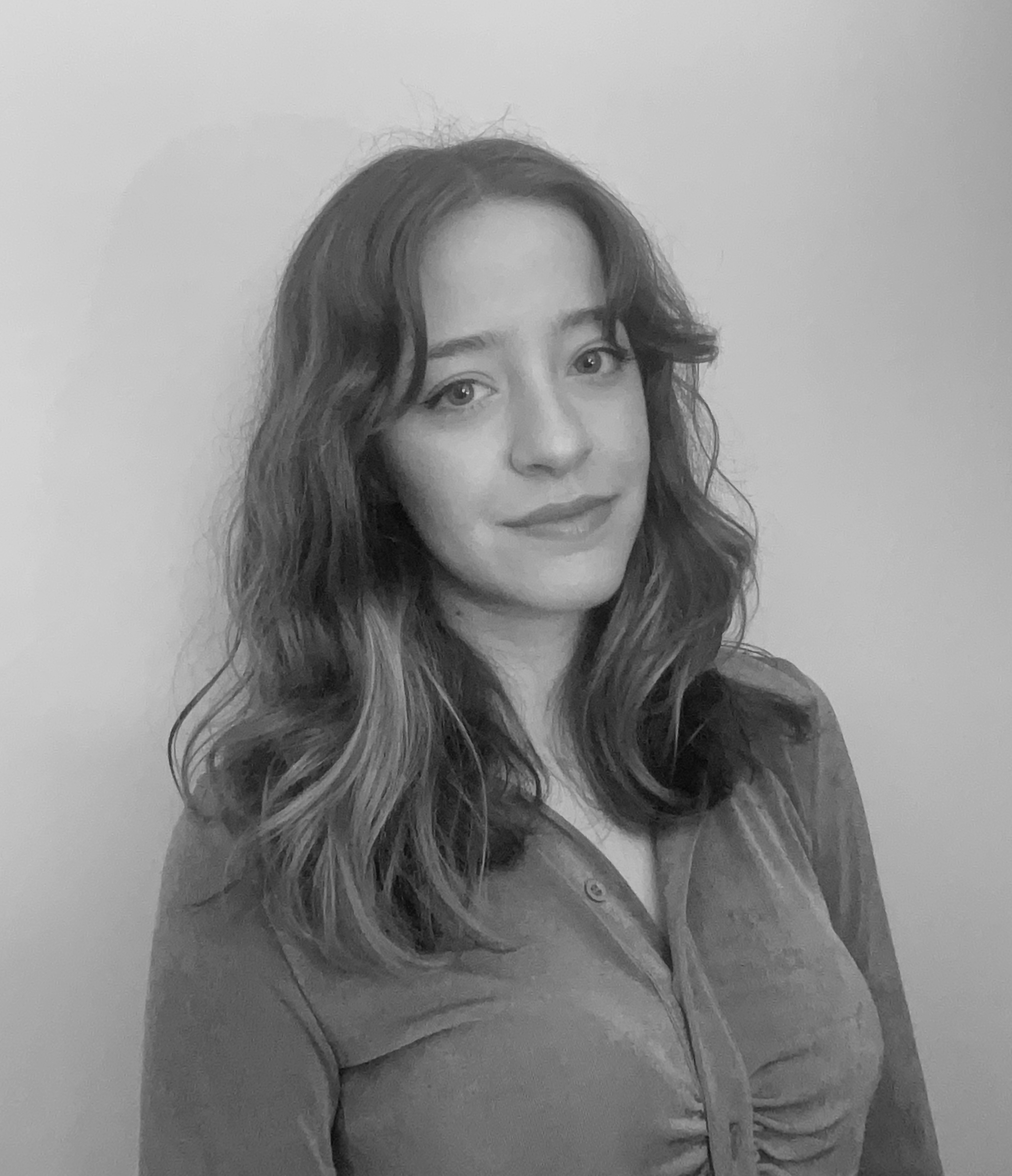}
\end{wrapfigure}\textbf{Elif Sena Kuru}
is an undergraduate student pursuing her Bachelors degree in Computer Engineering at Middle East Technical University (METU) since 2019. Her research interests include cloud security and machine learning.\hfill{}
\\
\hfill{}
\\
\hfill{}
\\
\hfill{}
\ITUpar

\begin{wrapfigure}{l}{0.32\columnwidth}
    \vspace{-.1in}
    \includegraphics[width=0.39\columnwidth]{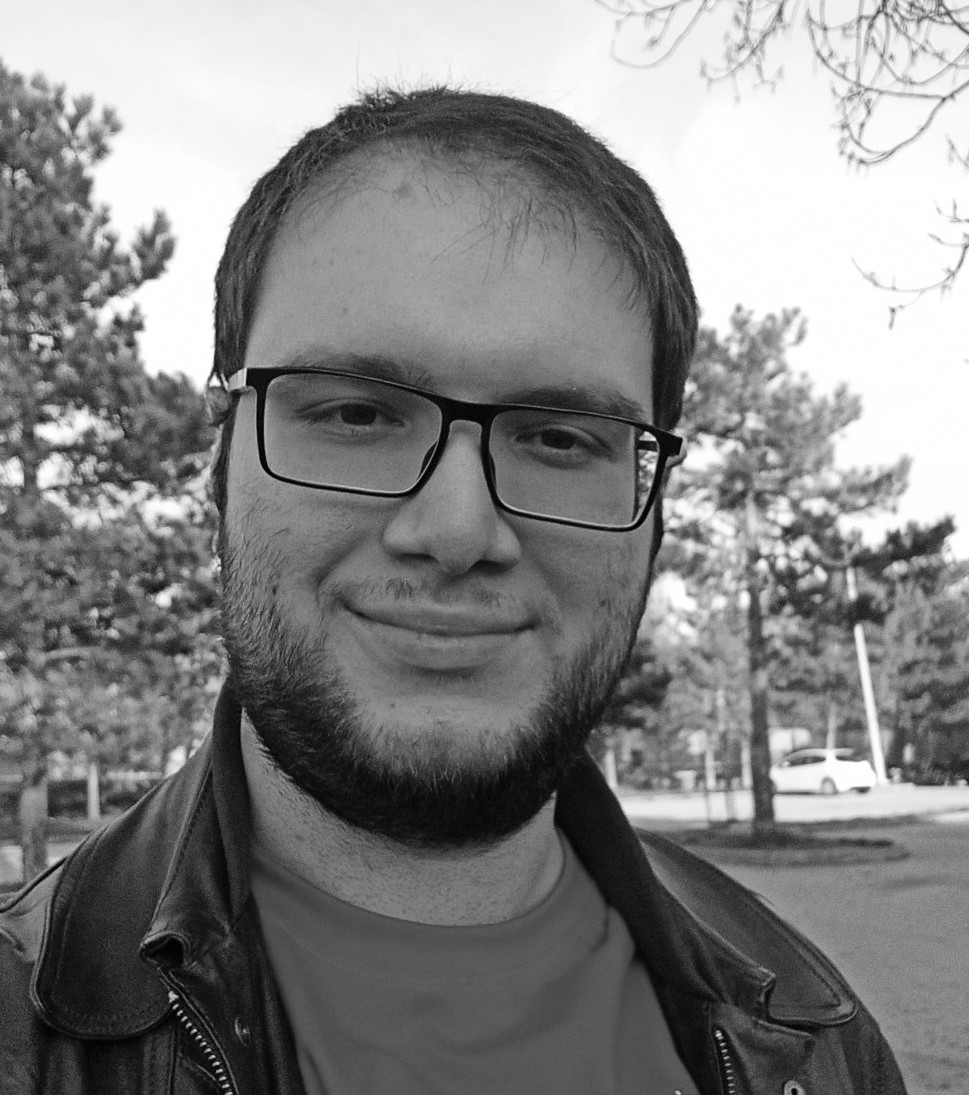}
\end{wrapfigure}\textbf{Yigit Sever}
received the B.S. degree in computer engineering at TED University, Turkey, in 2016, and the M.S. degree in computer engineering from Hacettepe University, Turkey, in 2019. He is currently a Ph.D. candidate in computer engineering at Middle East Technical University (METU), Turkey, where he is also working as a research assistant since 2020. His research interests include cloud security with a strong focus on container security, user and internet privacy and distributed systems. He is a member of Wireless Systems, Networks and Cybersecurity Laboratory, METU.\ITUpar

\begin{wrapfigure}{l}{0.32\columnwidth}
    \vspace{-.1in}
    \includegraphics[width=0.39\columnwidth]{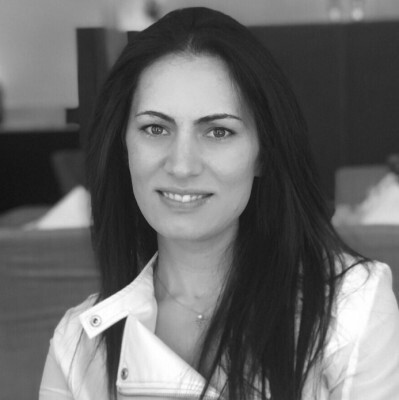}
\end{wrapfigure}\textbf{Pelin Angin} (Member, IEEE) received the B.S. degree in computer engineering at Bilkent University, in 2007, and the Ph.D. degree in computer science from Purdue University, USA, in 2013. From 2014 to 2016, she worked as a Visiting Assistant Professor and a Postdoctoral Researcher at Purdue University. She is currently an Assistant Professor in computer engineering at Middle East Technical University. Her research interests include the fields of cloud computing, the IoT security, distributed systems, 5G networks, data mining, and blockchain. She is among the founding members of the Systems Security Research Laboratory and an affiliate of the Wireless Systems, Networks and Cybersecurity Laboratory, METU.\ITUpar

\end{document}